\documentclass[letter]{aastex62}

\begin{document}

\title{Visualizing Multiwavelength Properties of Classified X-ray Sources from Chandra Source Catalog}

\correspondingauthor{Oleg Kargaltsev}
\email{kargaltsev@gwu.edu}

\author{Hui  Yang}
\altaffiliation{}
\affiliation{The George Washington University}

\author{Jeremy Hare}
\altaffiliation{}
\affiliation{NASA Goddard Space Flight Center}
\affiliation{NASA Post-doctoral Program Fellow}

\author{Igor Volkov}
\altaffiliation{}
\affiliation{The George Washington University}

\author{Oleg Kargaltsev}
\altaffiliation{}
\affiliation{The George Washington University}


\section{Abstract}

We present a simple but informative online tool to visualize the multiwavelength (MW) properties of $\approx2,700$ X-ray sources from Chandra Source Catalog version 2.0  with literature verified classifications. Here we describe the catalogs that we used to collect the classifications and extract the MW properties of these sources, and the properties themselves. We also describe the design and functionality of the tool.

\section{Background} 

X-ray sources can be classified in a number of different ways. For the brightest sources, one would typically study the X-ray spectrum (e.g., identifying Fe lines, fitting a set of models) and variability properties (e.g., flares, periodicity), in combination with any  multiwavelength (MW) counterpart properties.
Such comprehensive investigations become unfeasible for the much more numerous population of faint sources lacking high-quality X-ray spectra and  light curves. 
In such cases X-ray properties  are less certain and less informative while the MW counterpart properties (e.g., optical colors, X-ray to optical  flux ratios)  must be more heavily relied upon to classify a given source (see e.g., \citealt{2005ApJ...634L..53L,2008A&A...477..147C,2010ApJ...721.1663D,2012ApJ...756...27L,2012ApJ...745...99K,2017ApJ...837..130V,2018MNRAS.475.4841R,2020ApJ...889...53T}). 
Collecting the MW properties and being able to  visualize them is an important  first step.

Chandra Source Catalog  2.0\footnote{\url{https://cxc.cfa.harvard.edu/csc2/}} (\citealt{2020AAS...23515405E}; hereafter CSC 2.0) contains detailed information about positional, photometric, spectral, and temporal properties for 317,167 unique point and extended X-ray sources. 
 CSC 2.0 is the best catalog for collecting MW properties  of X-ray sources located in the Galactic plane due to the sub-arcsecond positional uncertainties\footnote{\url{https://cxc.cfa.harvard.edu/csc2/columns/positions.html}} (PUs) which result in a much lower chance of confusion when cross-matching. 

\section{Source sample}

We compiled X-ray sources belonging to 9 broad astrophysical classes: 
 active galactic nuclei (AGN), pulsars and isolated neutron stars (NS), non-accreting X-ray binaries (NS BIN)\footnote{These are wide-orbit binaries with millisecond pulsars, as well as red-back and black widow systems \citep{2019ApJ...872...42S}.}, cataclysmic variables (CV), high mass X-ray binaries (HMXB),  low-mass X-ray binaries (LMXB), high mass stars (HM-STAR)\footnote{These include Wolf-Rayet stars.}, low mass stars (LM-STAR) and young stellar objects (YSO).  The classifications were based on the following sources: Veron Catalog of Quasars $\&$ AGN 13th Edition \citep{2010A&A...518A..10V}, the ATNF Pulsar Catalog \citep{2005AJ....129.1993M}, the CV Catalog 2006 Edition \citep{2001PASP..113..764D}, the HMXB Catalog in the Galaxy 4th Edition \citep{2006A&A...455.1165L}, the LMXB Catalog 4th Edition \citep{ 2007A&A...469..807L}, the Catalogue of Stellar Spectral Classifications \citep{2014yCat....1.2023S},  the VIIth Catalog of Galactic WR Stars \citep{2001NewAR..45..135V}, the YSO catalogs from multiple  molecular clouds and open clusters \citep{2012AJ....144..192M,2011ApJS..194...14P,2005A&A...429..963O,2007A&A...463..275G,2011ApJS..196....4R,2011A&A...531A.141D}. 

We cross-matched sources from CSC 2.0  with these catalogs using error circles with the radius equal to the semi-major axis of the 2-$\sigma$ error ellipse in CSC 2.0. In our cross-matching, we avoided some crowded environments such as globular clusters and the Galactic center as well as sources strongly affected by complex extended emission around them (e.g., bright pulsar wind nebulae). 
Sources from populous classes (AGN, HM-STAR, LM-STAR and YSO) were  omitted if their X-ray PUs were $>1''$. There are several cases of X-ray detected sources from underpopulated classes that are offset by more than $1''$ from  their catalog positions,  likely due to poor absolute astrometry. For these sources, we confirmed the classifications/matches by reviewing the literature and inspecting the X-ray and MW images. Finally, we removed unreliable X-ray sources if they had NaNs and/or zero fluxes in multiple X-ray bands or if they had true {\em sat\_src\_flag} and/or {\em streak\_src\_flag} in CSC 2.0.  
  
The X-ray sources were then cross-matched, using X-ray error circles, to the optical Gaia eDR3 catalog \citep{2020arXiv201201533G}, the near-infrared Two Micron All Sky Survey \citep[2MASS;][]{2006AJ....131.1163S}, the CatWISE2020 catalog \citep{2021ApJS..253....8M}, the unWISE catalog \citep{2019ApJS..240...30S}, and the AllWISE catalog \citep{2014yCat.2328....0C} in the infrared.  
For X-ray sources having more than one MW counterpart located within the cross-matching radius, we took the nearest counterpart to the X-ray source as its MW counterpart. 
We  removed unreliable sources such as YSOs or STARs with no matched MW counterparts.
We also removed MW counterparts matched with isolated NSs by chance coincidence because virtually all of them are too faint to be detected in those surveys. As a result, we are left with 2,687 X-ray sources.  The number of sources for each source class is given at the top of Figure \ref{fig:1}. 

The visualization tool allows for plotting of various permutations of two of the following features:

\begin{itemize}
\item master-level X-ray properties\footnote{ described in detail in the CSC 2.0 website.} from CSC 2.0 including the  energy fluxes in the broad ($F_{\rm b}$; 0.5-7 keV), hard ($F_{\rm h}$; 2-7 keV), medium ($F_{\rm m}$; 1.2-2 keV), and soft ($F_{\rm s}$; 0.5-1.2 keV) bands, intra-observation Kuiper's test variability probability (P\_intra)\footnote{The conversion for both P\_intra and P\_inter shown in the plotting tool is designed to emphasize the extremely variable sources with variability probability larger than 0.9 where the most variable sources with probability of 1 correspond to the value of 3 after the conversion.}, inter-observation variability probability (P\_inter), and X-ray flux significance (Signif.);
\item three X-ray hardness ratios derived from the energy fluxes using HR$_{\rm ms}= (F_{\rm m}-F_{\rm s})/(F_{\rm m}+F_{\rm s})$, HR$_{\rm hm}= (F_{\rm h}-F_{\rm m})/(F_{\rm h}+F_{\rm m})$ and HR$_{\rm h(ms)}=(F_{\rm h}-(F_{\rm m}+F_{\rm s}))/(F_{\rm h}+F_{\rm m}+F_{\rm s})$; 
\item $G$, ${G_{\rm BP}}$ (BP) and ${G_{\rm RP}}$ (RP) magnitudes from the Gaia eDR3;
\item $J$, $H$, and $K$ magnitudes from the 2MASS;
\item $W1$ and $W2$ magnitudes from the CatWISE2020 and the unWISE and $W3$ magnitude from the AllWISE;
\item a selection of colors (e.g., $G-J$) using various MW magnitudes;
\item X-ray to optical flux ratio, $F_{\rm X}/F_{\rm o}$, with $F_{\rm X}=F_{\rm b}$ and $F_{\rm o}$ based on Gaia's $G$ band magnitude.
\end{itemize}

\section{The Visualization Tool GUI}

\begin{figure}[h!]
\begin{center}

\includegraphics[width=17cm,angle=0]{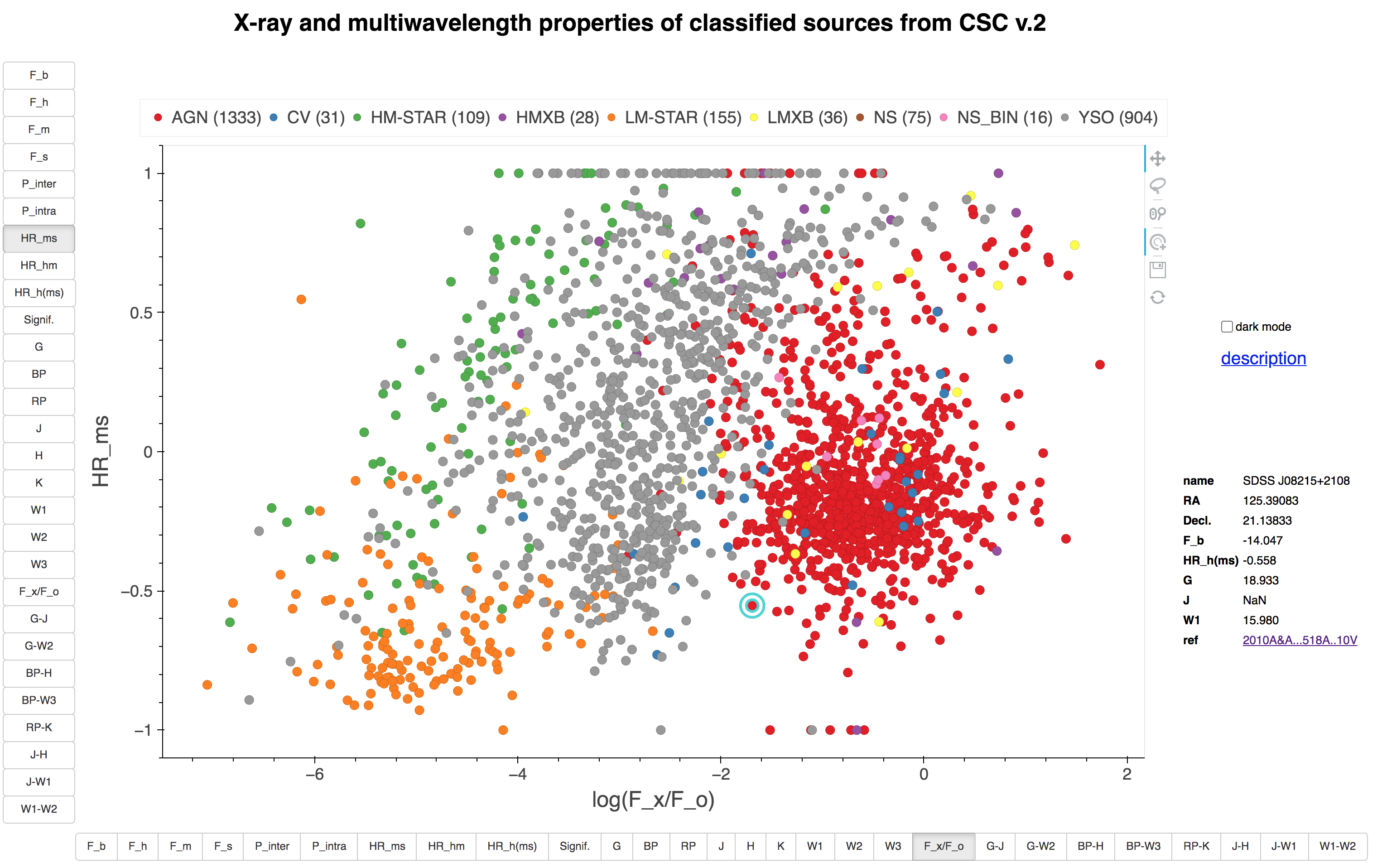}
 \includegraphics[width=17cm,angle=0]{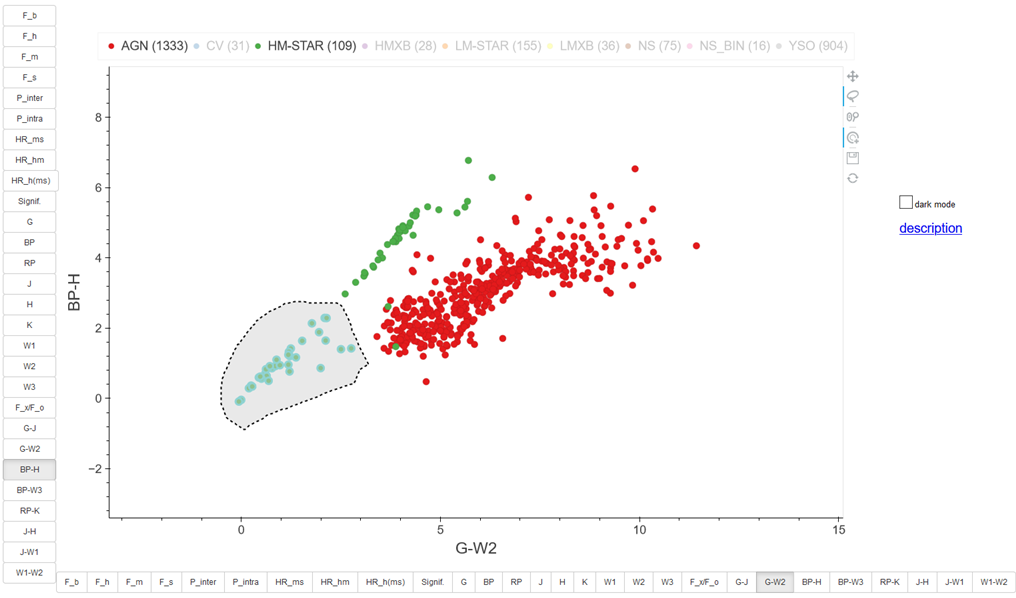}
\caption{
The upper panel shows a screenshot from the interactive visualization tool, available at \url{https://home.gwu.edu/~kargaltsev/XCLASS/},of HR$_{\rm ms}$ hardness ratio versus X-ray to optical flux ratio for all nine classes of sources. The lower panel shows another screenshot of $BP$-$H$ color versus $G$-$W2$ color for AGNs in red and HM-STARs in green while a group of HM-STARs (within the shaded area) is selected by the {\em lasso} tool. 
\label{fig:1}}
\end{center}
\end{figure}

The tool, publicly available on a dedicated webpage, \url{https://home.gwu.edu/~kargaltsev/XCLASS/}, allows for  easy  interactive plotting of MW properties of the sources for a user-selected subset of source classes.  The  Graphical User Interface (GUI) is written in Java while plotting is done on-the-fly using  the Bokeh Python library\footnote{\url{https://bokeh.org}}.  As shown in Figure \ref{fig:1}, the user can choose from multiple MW features by clicking the corresponding buttons to the left of the Y-axis and below the X-axis. At the top there are buttons corresponding to the distinct astrophysical classes with the numbers of sources within each class. Users can  choose to display different combinations of the classes by toggling/untoggling the class names. Whenever appropriate, a logarithmic scale is used in the plot (which is reflected in the axis labels). The user  can select a source by clicking on it and some information about the source will be shown.  A {\em lasso} tool can be used to select a group of sources which then remain selected for any combination of features plotted.   The plot can be easily saved as a PNG file by clicking the corresponding button to the right of the plot.     

An example, with HR$_{\rm ms}$ hardness ratio plotted versus X-ray to optical flux ratio,  is shown in the upper panel of Figure \ref{fig:1}.  One can see a clear separation between AGN and HM-STAR/LM-STAR/YSO while other types of sources are overlapping, requiring an investigation of different sets of features to distinguish between them. Another example, in the lower panel if Figure \ref{fig:1}, shows the $BP$-$H$ color against $G$-$W2$ color for AGN in red and HM-STAR in green. A subset of HM-STAR with bluer colors compared to other HM-STAR and AGN is selected by the {\em lasso} tool marked by highlighted cyan circles and shaded region will remain highlighted after changing to other combination of features plotted.

\section{Summary}
We provide an online tool, based on previously published X-ray source classifications and data from  CSC 2.0 and multiple  all-sky surveys, 
 to visualize  MW properties of X-ray sources with known classes. We ask anyone who will be using this tool in their work  to cite this publication.

\acknowledgments

Support  for  this  work  was  provided  by NASA through CXO Awards AR8-19009B, AR9-20005A,  AR0-21007X, and  NASA ADAP award 80NSSC19K0576. 
JH acknowledges support from an appointment to the NASA Postdoctoral Program at the Goddard Space Flight Center, administered by the USRA through a contract with NASA.

\end{document}